\documentclass[conference]{IEEEtran}
\IEEEoverridecommandlockouts

\usepackage{cite}
\usepackage{amsmath,amssymb,amsfonts}
\usepackage{algorithmic}
\usepackage{graphicx}
\usepackage{textcomp}
\usepackage{multirow}
\usepackage{xcolor}
\usepackage{xspace}
\usepackage{tabularray}
\usepackage{booktabs}
\usepackage{siunitx}
\usepackage{color,soul}
\usepackage{url}
\usepackage{doi}
\def\BibTeX{{\rm B\kern-.05em{\sc i\kern-.025em b}\kern-.08em
    T\kern-.1667em\lower.7ex\hbox{E}\kern-.125emX}}

\newcommand\CR[1]{\ensuremath{\mathit{C\kern-0.3exR}_{\mathit{#1}}}}
\newcommand\CC{\ensuremath{\mathit{C\kern-0.3exC}}\xspace}
\newcommand\RRMSE{\ensuremath{\mathit{R\kern-0.3exR\kern-0.3exM\kern-0.3exS\kern-0.3exE}}\xspace}

\DeclareMathOperator*{\cov}{Cov}
    
\begin{document}

\title{PhysioEdge: Multimodal Compressive Sensing Platform for Wearable Health Monitoring\\
\thanks{This work has been supported by the Flanders Chips Competence Center (FC3). FC3 is co-funded by the European Union. The project is supported by the CHIPS JU and its members (including top-up funding by VLAIO).}
}

\author{\IEEEauthorblockN{
Rens Baeyens\IEEEauthorrefmark{1}\IEEEauthorrefmark{2},
Dennis Laurijssen\IEEEauthorrefmark{1}\IEEEauthorrefmark{2},
Jan Steckel\IEEEauthorrefmark{1}\IEEEauthorrefmark{2} and
Walter Daems\IEEEauthorrefmark{1}\IEEEauthorrefmark{2}}
\IEEEauthorblockA{\IEEEauthorrefmark{1}FTI Cosys-Lab,
University of Antwerp,
Antwerp, Belgium, Email: walter.daems@uantwerpen.be}
\IEEEauthorblockA{\IEEEauthorrefmark{2}Flanders Make Strategic Research Centre,
Lommel, Belgium}}
\maketitle

\begin{abstract}
The integration of compressive sensing with real-time embedded systems opens new possibilities for efficient, low-power biomedical signal acquisition. This paper presents a custom hardware platform based on the RP2350 micro-controller, tailored for synchronized multi-modal biomedical monitoring. The system is capable of capturing cardiopulmonary sounds, along with biopotential signals such as phonocardiography (PCG), electrocardiography (ECG) and electromyography (EMG), photoplethysmography (PPG), and inertial measurement unit (IMU) data for posture recognition. To ensure sample-accurate synchronization, a Sub-1GHz radio system is used across multiple nodes. Wi-Fi and Bluetooth connectivity enable centralized data aggregation. Experimental results demonstrate the achieved decrease in power consumption when using compressive sensing, efficient multi-node synchronization, and scalability for wireless biomedical monitoring applications. The compact form factor and low-cost design make it suitable for various medical applications, including remote healthcare and long-term monitoring.
\end{abstract}

\begin{IEEEkeywords}
Biomedical Monitoring, Compressive Sensing, Embedded Systems, Synchronization
\end{IEEEkeywords}

\section{Introduction}
Biomedical signal monitoring is a cornerstone of modern healthcare, enabling continuous tracking of physiological parameters for early diagnosis and management of chronic conditions. Wearable devices for biomedical monitoring have become widely prevalent, offering real-time health insights and facilitating remote patient care. However, traditional high-resolution sampling methods generate large volumes of data, straining bandwidth and power resources, particularly in portable and remote settings. Although various low-power strategies have been explored, compressive sensing has not been sufficiently leveraged as a means to reduce power consumption in these devices. 

The implementation of multi-modal biomedical monitoring on edge devices brings forth several key challenges \cite{Iadarola2024}. Traditionally, high-fidelity data requires high data rates, whereas embedded devices typically pose bandwidth constraints. Additionally, edge devices are typically battery-powered, necessitating the adoption of low-power solutions. These solutions can take many forms \cite{Cho2017,Biagetti2021}, with compressive sensing (CS) emerging as a promising approach to mitigate data throughput and power consumption challenges \cite{Chen2024}. CS leverages signal sparsity to reduce sampling rates without significantly compromising signal quality, making it a compelling solution for biomedical signal acquisition. Its application to physiological signals\textemdash such as cardiopulmonary sounds, electrocardiography, and pulse oximetry\textemdash can drastically reduce data acquisition and transmission overhead while maintaining diagnostic integrity.

To ensure usability of data from different sensing nodes, synchronization across multiple nodes must be achieved \cite{Djenouri2016,Coviello2020,Sondej2024,Biagetti2025,Derogarian2014}. Section~\ref{methods} presents the custom biomedical acquisition hardware based on an RP2350 micro-controller, which enables real-time, multi-modal signal acquisition, including PCG, ECG, EMG, inertial measurement, and PPG signals. Synchronization is achieved using a CC1120 $\qty{868}{M\hertz}$ radio module, while data transmission occurs over Wi-Fi or Bluetooth. The signals acquired using CS are reconstructed at a central node using a Convolutional Neural Network (CNN) for CS signal reconstruction, previously shown to achieve high-fidelity reconstructions for cardiopulmonary sounds \cite{Baeyens2025}. The proposed hardware platform is evaluated in terms of data integrity, power efficiency, and synchronization accuracy, demonstrating its potential for energy-efficient, real-time biomedical monitoring in portable and wearable applications, with detailed results provided in Section~\ref{results}.

\begin{figure*}[!ht]
\centering
  \centering
  \includegraphics[width=0.97\textwidth]{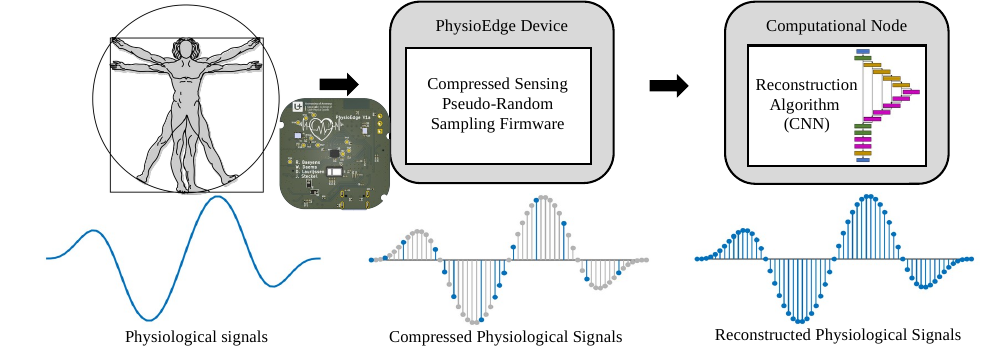}%
  \caption{Overview of the proposed system. The physiological sounds are first recorded using the PhysioEdge device with pseudo-random
sampling firmware for compressed sensing. Second, the compressed signals are sent to a computational node that executes the CNN reconstruction algorithm
to obtain the reconstructed cardiopulmonary sounds. This paper focuses specifically on the PhysioEdge Device.}
  \label{fig:system_overview}
\end{figure*}

\section{Methods}
\label{methods}
 \subsection{Hardware Design}
\label{hardware_design}
At the core of the PhysioEdge Compressive Sensing Node is the RP2350B micro-controller from Raspberry Pi. Wireless communication is managed via a CYW4343W module from Cypress, supporting both Bluetooth and Wi-Fi connectivity. A Sub-1GHz synchronization channel is implemented using the CC1120 radio from Texas Instruments. The RP2350B features dual ARM Cortex-M33 processors, dual Hazard3 RISC-V processors, a programmable I/O (PIO) subsystem, and multiple digital interfaces, providing high design flexibility.\\
For acoustic cardiopulmonary signal acquisition, two SPH0645LM4H-B MEMS microphones are utilized to sample acoustic signals in stereo at either \qty{4}{\kilo\hertz} or \qty{8}{\kilo\hertz} depending on the use case. 
For heart sounds, i.e. phonocardiogram (PCG) signals, there is a lot of variation in the literature discussing the frequency range. The main difference lies in whether heart murmur is taken into account or not, since this is typically situated at a higher frequency than the systole and diastole cycles \cite{Chowdhury2020,Ghosh2022,Behera2024, Varshney2020}. For this reason, the sampling frequency for PCG acquisition is set to \qty{4}{\kilo\hertz}, to cover all potential use-cases, this matches the sampling frequency used in the CirCor DigiScope Dataset \cite{Oliveira2022}. For respiratory sounds, the useful frequency range extends up to \qty{4}{\kilo\hertz}, leading to a sample frequency of \qty{8}{\kilo\hertz} \cite{Kahya2018}. \\
For biopotential signal acquisition, an ST1VAFE3BX analog front end is implemented. This chip, designed for ECG, EEG and EMG measurements, features low power consumption, programmable gain, 12-bit resolution, and a sampling rate of up to \qty{3.2}{\kilo\hertz}, ensuring flexiblity for a wide variety of applications \cite{Crippa2024}.\\
To capture motion data, the system integrates a BMI323 IMU sensor from Bosch SensorTec. This low-power 3-axis accelerometer and gyroscope offers high-precision motion tracking, with sensitivities of 0.06 mg for acceleration and 0.004 dps for angular velocity, at a maximum sample rate of \qty{6.4}{\kilo\hertz}.\\
For photoplethysmography (PPG) measurements, the system employs the MAX30102 PPG sensor from Maxim Integrated. This sensor is specifically designed for advanced health monitoring applications, enabling high-resolution heart rate and blood oxygen saturation measurements while maintaining minimal power consumption.\\
A synchronization mechanism is implemented using a Sub-1GHz communication channel with a CC1120 radio transceiver from Texas Instruments. This module features a wave-matching correlation filter that detects the synchronization word at the start of a packet and triggers a digital pin. By processing synchronization triggers without requiring full packet parsing, this method significantly enhances synchronization accuracy. \\
Finally, all acquired data is transmitted to a central acquisition node using either Wi-Fi or Bluetooth, facilitated by a CYW4343W radio module.

\subsection{Firmware}
\subsubsection{Data Acquisition}
The custom firmware is developed using mainly the C SDK from Raspberry Pi, which is well documented. Due to the relatively low sample frequencies, the data acquisition of the BMI323, ST1VAFE3BX and MAX30102 are handled on an interrupt basis. The I2S SPH0645LM4H-B stereo setup requires the generation of a data clock running between $1$ and $\qty{4}{M\hertz}$, as well as a digital read on every low and high edge of the data clock. This was implemented in a PIO assembly program to ensure uninterrupted sampling. The data is then transferred from the PIO receive FIFO buffer to a dual buffer in main system memory using a DMA channel. The DMA channel then triggers an interrupt each time one of the buffers is full to signal that it is ready for transmission. If compressive sensing is enabled, the main core then applies pseudo-random decimation according to the desired compression ratio \CR{} on the data as described in the following Section~\ref{cs}. Implementation of the pseudo-random decimation in the PIO module instead of the main core is future work.

\begin{figure}[ht]
\centering
  \centering
  \includegraphics[width=\columnwidth]{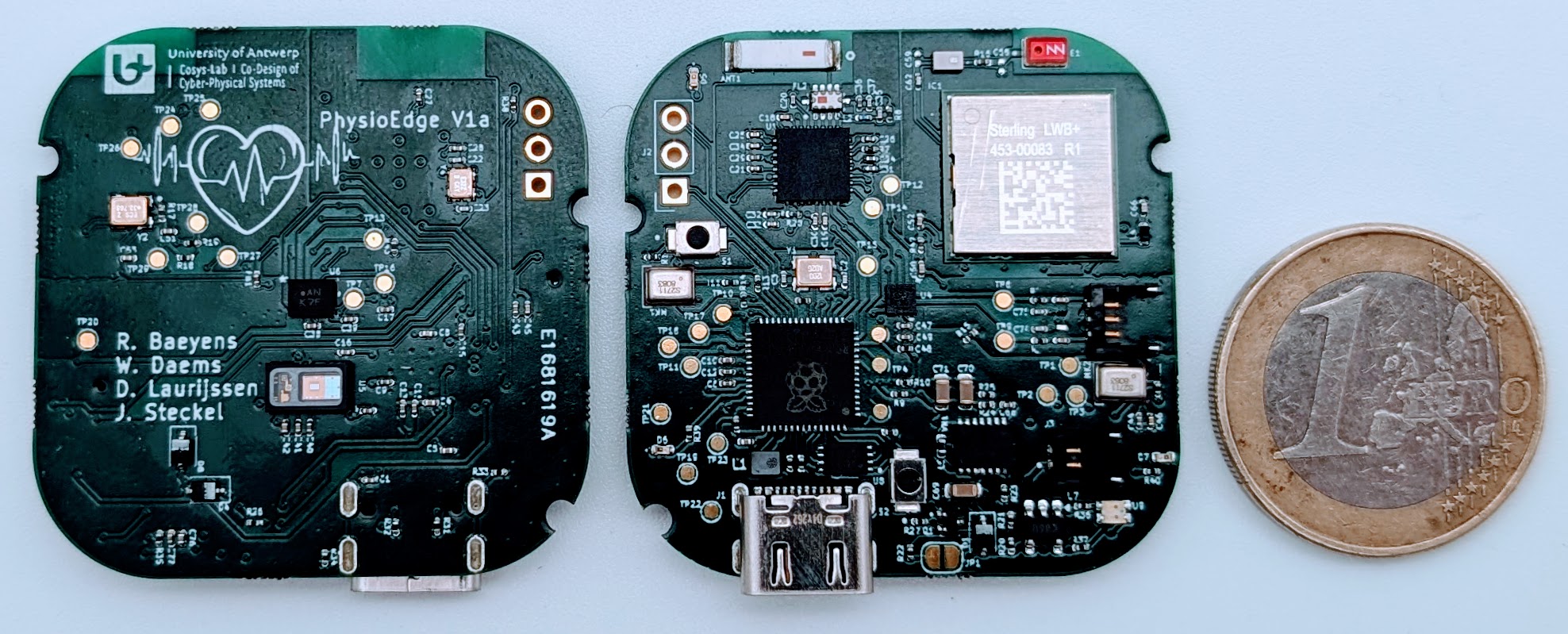}
  \includegraphics[width=\columnwidth]{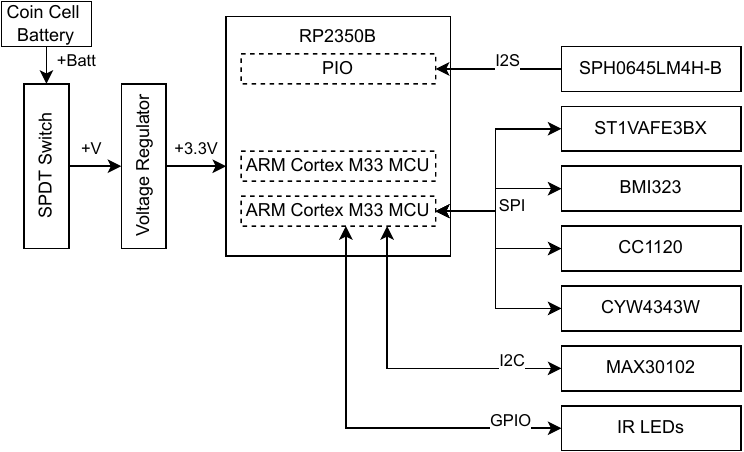}%
  \caption{Developed PCB and system architecture of the PhysioEdge Compressive Sensing Node with the RP2350B at its core. Two SPH0645LM4H-B are sampled using a PIO module and transferred using a DMA channel. The ST1VAFE3BX biopotential and BMI323 IMU sensing modalities communicate over an SPI bus that is shared with the CC1120 Sub-1GHZ transceiver and the CYW4343W radio module. The MAX30102 PPG sensor communicates over I2C. IR LEDs are provided for synchronization and localization with an infrared motion capture system. }
  \label{fig:hardware_arch}
\end{figure}

\subsubsection{Compressive Sensing}
\label{cs}
Compressive sensing (CS) enables efficient signal acquisition by exploiting sparsity to reconstruct signals from fewer samples than required by the Nyquist criterion \cite{Candes2008}. In the proposed sensing node for physiological signal acquisition, we implement a pseudo-random moving decimation (PRMD) approach to reduce data while maintaining reconstruction accuracy \cite{OConnor2014}. Rather than applying a fixed decimation pattern, which risks aliasing and structured loss of information, we use an XORShift pseudo-random number generator (PRNG) to dynamically determine the retained samples \cite{Marsaglia2003}. XORShift is chosen for its computational efficiency, low memory footprint, and statistically adequate randomness, making it well-suited for embedded implementations. This method ensures that sampling remains sufficiently incoherent with respect to the signal basis, preserving essential information for CS reconstruction while minimizing computational overhead.
The amount of samples $d$ to skip is calculated for each subsequent sample in the following manner: 
\begin{equation}
\label{eqn:modulo}
d \equiv x\!\!\!\!\mod(2\cdot\CR{}),\text{~~with~}x\in \{ 0,1,...,2^{32}-1\}
\end{equation}
with $x$ a pseudo-random number generated using the XORShift32 algorithm, achieving a periodicity of $2^{32}-1$ \cite{Marsaglia2003}. This pseudo-random number is then modulo divided by twice the compression ratio, to result in a total average step size equal to the compression ratio maintaining a uniform sampling delay distribution on the interval $1$ to $2\cdot\CR{}$.

\subsection{CNN-based Reconstruction}
The main goal of CS reconstruction algorithms is to reconstruct an estimate $\tilde{s}[k]$ of the signal $s[k]$ from the randomly sub-sampled version of this signal $s_c[k_r]$. The typical approach to reconstructing $\tilde{s}[k]$ is to solve the sparse reconstruction problem, which can be modeled as: 
\begin{equation}
\label{eq:fullcs}
    s_c[k_r] = \Phi_r \cdot \Psi_b \cdot D \cdot x + \eta
\end{equation}
where $s_c$ contains noisy CS measurements, $\Phi_r$ is the random sampling matrix, $\Psi_b$ is a basis function transform from time domain into a domain where $s[k]$ is considered sparse, $D$ is a dictionary with the basis functions in the sparse basis for $s[k]$, $x$ is a signal with a maximum of $K$ non-zero elements and is therefore a so-called $K$-sparse signal, and $\eta$ is used to model Gaussian random noise. The reconstruction can then be rewritten to: 
\begin{equation}
\label{eq:reconstruction}
    \tilde{s}[k] = \Psi_b \cdot D \cdot x
\end{equation}
Previous research \cite{Baeyens2025} has demonstrated that a Convolutional Neural Network (CNN)-based approach for CS reconstruction can achieve unprecedented reconstruction accuracy compared to traditional CS reconstruction algorithms such as Compressive Sampling Matching Pursuit (CoSAMP) and Optimized Discrete Cosine Transform (ODCT) \cite{Needell2009, Chen2024}.
The CNN-based reconstruction algorithm relies on a pseudo-random division of the compressed signal into embedding vectors to increase the incoherence. Before the embedding signals $s_c^n [k_{r}]$ are sent to the CNN for reconstruction, they are first linearly interpolated to $s_b^n [k_{i}]$ so that they are sampled regularly and have a fixed length of 34,976 samples. This interpolation operation introduces artifacts in the frequency spectra of $S_b^n$, which are resolved by the CNN.
In practice, these embedding vectors are different under-sampled versions of the same signal. Because we are taking multiple independent pseudo-random embedding vectors from the same signal, certain samples from the original signal may be used multiple times by separate embeddings.

Mathematically, a neural operator $G$ is applied to find the reconstruction $\tilde{s}[k]$ from $N$ interpolated embedding vectors $s_{b}^{n}[k_i]$.
\begin{equation}
    \tilde{s}[k]= G \left( s_b^1 [k_{i}], \dots, s_b^N [k_{i}] \right)
\end{equation}
Unlike conventional iterative reconstruction techniques, which often suffer from high computational complexity and require careful parameter tuning, CNN-based reconstruction leverages deep learning to directly learn an optimal mapping from compressed measurements to high-fidelity signal representations. This data-driven approach enables faster and more accurate reconstructions while effectively capturing complex, non-linear structures in biomedical signals. In this work, we employ the CNN architecture from \cite{Baeyens2025}, trained separately on the SPRSound respiratory sounds dataset and the Circor PCG dataset \cite{Zhang2022, Oliveira2022}. 
\subsection{Synchronization Mechanism}
\label{synchronization}
\begin{figure}[!ht]
\centering
  \centering
  \includegraphics[width=0.9\columnwidth]{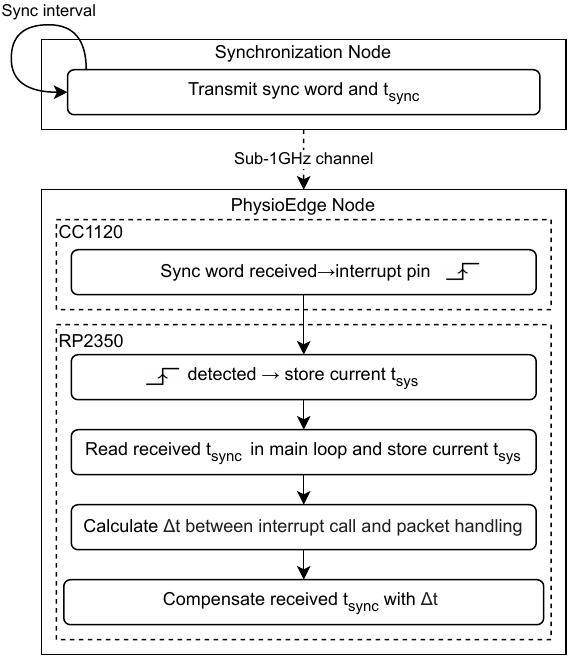}
  \caption{Schematic of the synchronization mechanism.}
  \label{fig:synchronization_mechanism}
\end{figure}

A synchronization mechanism is continuously active. The system clock comes from a \qty{12}{\mega\hertz} crystal oscillator with a frequency stability of $\pm$\qty{10}{ppm}. With the RP2350 running at \qty{150}{\mega\hertz}, this implies a maximal clock drift of \qty{10}{\micro\second} per second. To ensure single-sample accuracy with a strong margin, the synchronization message is transmitted every five seconds. It is illustrated in Figure~\ref{fig:synchronization_mechanism}. This synchronization message contains a synchronization word and a timestamp of the synchronization node. When the synchronization word is detected in the CC1120 receiver's correlation filter, an interrupt is generated. When this interrupt is handled, the current system time of the edge device is stored. The incoming synchronization packet is then handled at a suitable time, with the only requirement being that the packet is parsed before the next synchronization message arrives. When the synchronization packet is actually parsed, the time stored at interrupt detection is subtracted from the current system time to account for the delay between packet detection and packet handling. This adjusted timestamp is then embedded in the sensor data, allowing full synchronization between the sensor node network. The achieved synchronization accuracy is presented in Section~\ref{results}. 

\section{Results}
\label{results}
\subsection{Evaluation Metrics}
\begin{figure}[t]
\centering
  \centering
  \includegraphics[width=\columnwidth]{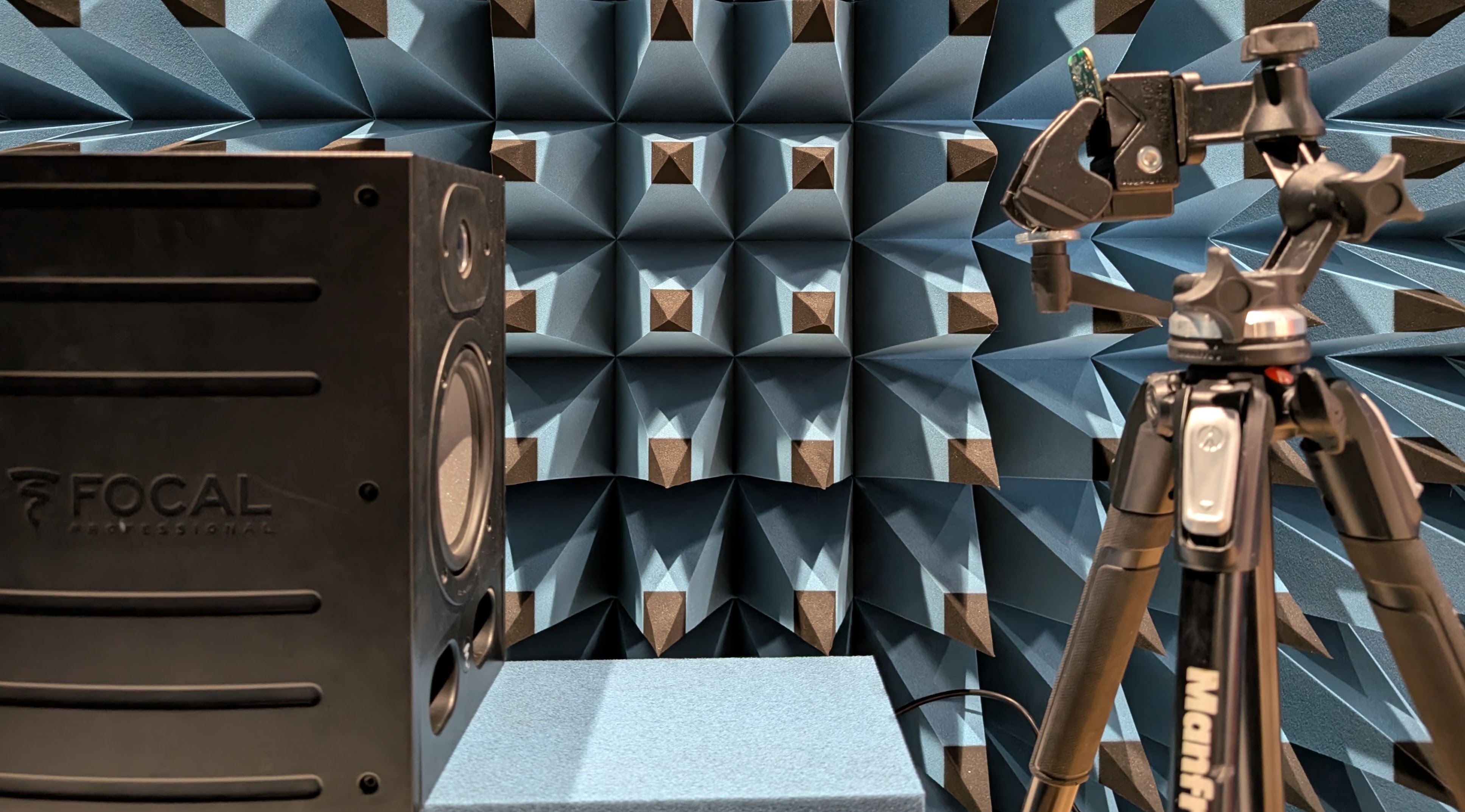}%
  \caption{Setup used to evaluate the microphone signal integrity when applying CS in real-time on an embedded device. A studio monitor was used to playback lung sounds to be recorded with the SPH0645LM4H-B using compressive sensing.}
  \label{fig:recording_setup}
\end{figure}

The proposed system is evaluated across three key performance metrics: power consumption, synchronization accuracy, and signal integrity. These metrics assess the system’s improvements and highlight potential trade-offs. The primary advantage of applying CS is the reduction in the number of samples acquired and transmitted, leading to lower power consumption. To quantify this effect, power measurements were conducted under various scenarios using a Nordic Power Profiler Kit II. The accuracy of synchronization is assessed by measuring the maximum time difference between two synchronization receiver nodes in response to a synchronization interrupt. 
To validate the signal integrity, an alternative approach is required since obtaining a ground truth measurement for lung sounds in a controlled manner is challenging. We utilized a high-quality studio speaker with a well defined frequency response to playback a subset of pre-recorded lung sounds from the SPRSound lung sound dataset \cite{Zhang2022} in an anechoic chamber as shown in Figure~\ref{fig:recording_setup}. This chamber is not specifically designed for acoustic anechoic purposes, but it was the most ideal room available to us to perform this experiment. These sounds were then recorded using the proposed compressive sensing method with one of the SPH0645LM4H-B microphones of the PhysioEdge. This setup ensured that the input to our system was well-characterized and reproducible, allowing us to assess the fidelity of reconstructed signals under varying compression ratios. 
These signals are compressed using the same compression ratios as in a simulation-based study and reconstructed using the same CNN model \cite{Baeyens2025}. The resulting reconstructions are then compared with the results achieved in simulation.
\begin{table}[b]
\centering
\caption{Measured Power consumption During PCG signal transmission}
\begin{tabular}{cccc} 
\toprule
  Method & \CR{} & Datarate $\downarrow$ & Average Power $\downarrow$\\ 
 \midrule
  Wi\textendash Fi & / & 1 Mbps & 25.0 mW  \\ 
  Wi\textendash Fi & 10 & 100 kbps & 23.5 mW \\
  Wi\textendash Fi & 30 & 33 kbps & 22.8 mW \\ 
  Bluetooth & 10 & 100 kbps & 6.6 mW \\
 
 Bluetooth & 30 & 33 kbps & 4.9 mW \\ 
 \bottomrule
\end{tabular}
\label{table:powermeasurements}
\end{table}

\begin{table}[!b]
\centering
\caption{CS-CNN Reconstruction Evaluation}
\begin{tabular}{cccc} 
 \toprule
    & \CR{} & \CC & \RRMSE \\ 
 \midrule
 \multirow{3}{*}{Simulated CS} & 4 & 0.96 $\pm$ 0.01  & 0.27 $\pm$ 0.01 \\
    & 10 & 0.94 $\pm$ 0.01 & 0.35 $\pm$ 0.03 \\
    & 30 & 0.63 $\pm$ 0.19 & 0.76 $\pm$ 0.13\\
 \cmidrule(lr){2-4}
 \multirow{3}{*}{Embedded Real-time CS} & 4 & 0.96 $\pm$ 0.01  & 0.28 $\pm$ 0.02 \\ 
     & 10 & 0.93 $\pm$ 0.02 & 0.36 $\pm$ 0.04 \\
     & 30 & 0.65 $\pm$ 0.18 & 0.74 $\pm$ 0.12\\
 \bottomrule
\end{tabular}
\label{table:comparison}
\end{table}

\subsubsection{Power Consumption}
\label{powerconsumption}
To assess the impact of CS on power consumption, five experimental scenarios were evaluated. As a baseline, the system samples and transmits uncompressed signals from two microphones over Wi-Fi at a data rate of 1 Mbps\textemdash approaching the theoretical maximum throughput for Bluetooth LE. To compare this baseline with CS-enabled transmission, signals were compressed at ratios of 10 and 30, reducing the data rate to 100 kbps and 33 kbps, respectively. The compressed data was transmitted over both Bluetooth and Wi-Fi. Table~\ref{table:powermeasurements} summarizes the measured power consumption for each configuration.

\begin{figure*}[!ht]
\centering
  \centering
  \includegraphics[width=\textwidth]{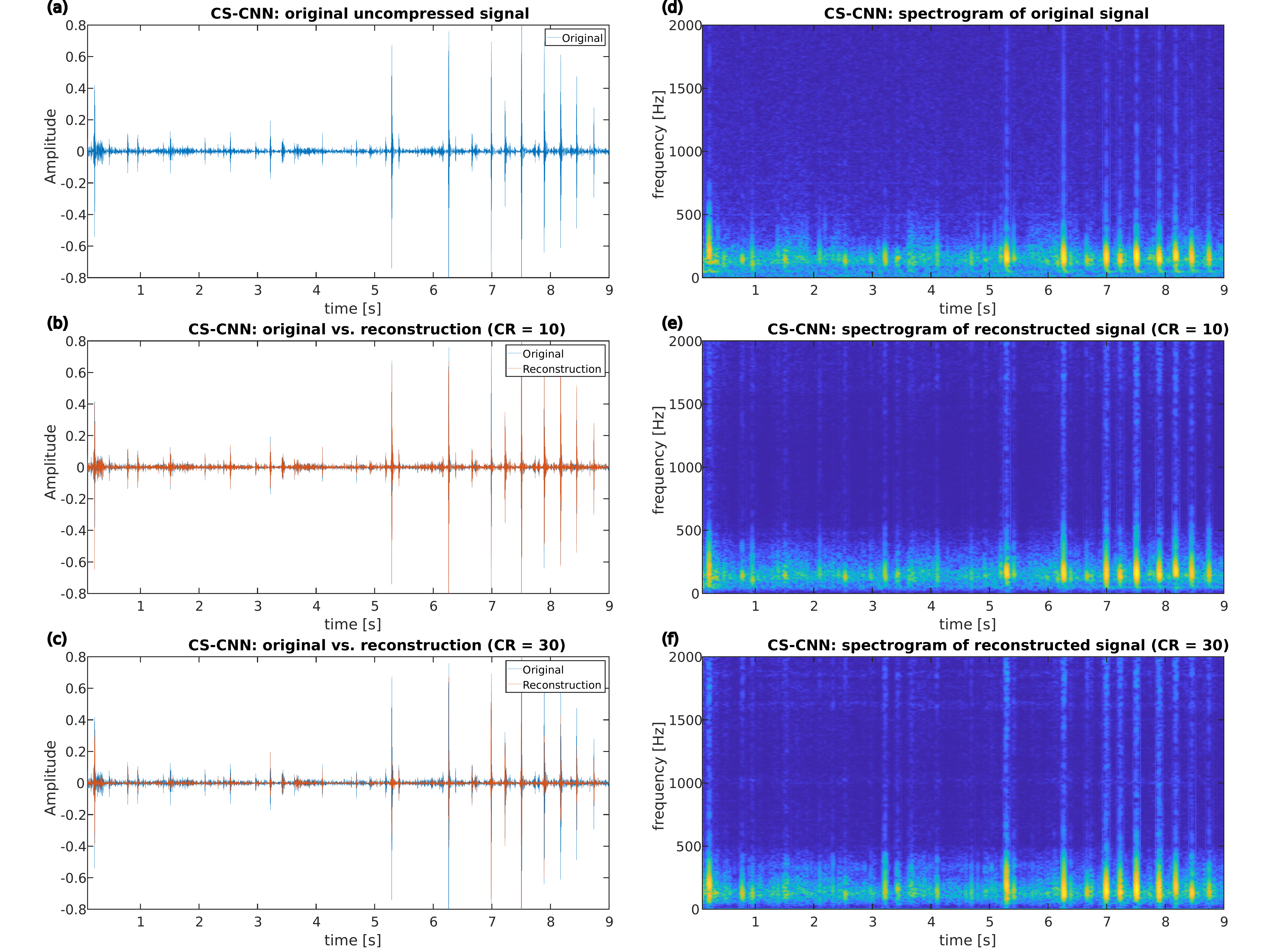}%
  \caption{Comparison between fully sampled data and its reconstruction using the CS-CNN approach with Real-time CS on the PhysioEdge using a compression ratio of 10 and 30. (a) The original signal in the time domain. (b) The original signal in comparison with the reconstruction using CR = 10 in the time domain. (c) The original signal in comparison with the reconstruction using CR = 30 in the time domain. (d) Presents the spectrogram within the region of interest for respiratory sound analysis of the original signal. (e) Presents the spectrogram of the reconstructed signal using a compression ratio of 10. (f) Presents the spectrogram of the reconstructed signal using a compression ratio of 30.}
  \label{fig:reconstruction_results}
\end{figure*}

\subsubsection{Multi-Node Synchronization Accuracy}
To validate the proposed synchronization mechanism, two edge nodes and a central timekeeper node are used. The central node transmits its own system time every 5 seconds over the Sub-1GHz channel. Receiving this timestamp triggers an interrupt at the edge device. The time difference in sync word detection by the CC1120 radio transceiver is measured by sampling the CC1120 sync detect interrupt pin. The interrupt handling time between the two edge nodes is then detected by toggling a pin in the interrupt routine on both devices. This pin toggle is placed at the same point where the system time is stored during normal operation. This response time difference was measured over a period of 10 minutes using a Saleae Logic Pro, sampling at \qty{500}{M\hertz}, to evaluate the stability and deviation in response time. A median response time difference of $\qty{2,65}{\micro\second}$ with a standard deviation of $\qty{2,06}{\micro\second}$ was recorded. The maximum recorded time difference between the response time of the nodes was $\qty{11.1}{\micro\second}$. Since the highest sampling frequency of the sensing modalities on the edge device is \qty{8}{\kilo\hertz}, this implies that the synchronization is single-sample accurate for all implemented modalities. In fact the synchronization mechanism provides single-sample accuracy for sampling frequencies up to $\qty{90.09}{\kilo\hertz}$. 

\subsubsection{Signal Integrity}
To validate the signal integrity of the acquired data using compressive sensing on an embedded device, lung sounds were captured at a sampling rate of 8 kHz with three different compression ratios: 4, 10, and 30. These compression settings were selected to match those used in a simulation-based study found in \cite{Baeyens2025}, enabling a direct comparison between simulated and real-world implementations.

The reconstruction results are compared with the results achieved in simulation on the selected subset. The following evaluation metrics are used: Compression Ratio (\CR{}), Relative Root Mean Square Error (\RRMSE), and Pearson's Correlation Coefficient (\CC). These metrics are defined as:
\begin{equation}
    \CR{} = \frac{\text{Size of initial signal}}{\text{Size of compressed signal}}
\end{equation}
\begin{equation}
    \RRMSE = \sqrt{\frac{\sum\limits_{k}(\tilde{s}[k]-s[k])^{2}}{\sum\limits_k s^{2}[k]}}
\end{equation}
\begin{equation}
    \CC = \frac{\cov(s[k],\tilde{s}[k])}{\sigma(s[k])\sigma(\tilde{s}[k])}
\end{equation}
The results for these metrics are presented in Table~\ref{table:comparison}.

An example resulting from these reconstructions is presented in Figure~\ref{fig:reconstruction_results} to demonstrate the feasibility of real-time compressive sensing for lung sound acquisition in low-cost embedded systems. The results expose the trade-off between reconstruction quality and compression ratio. It is noteworthy that even for a compression ratio of 30, the trend of the signal is maintained. However, further evaluation is necessary to validate the clinical applicability of higher compression ratios towards e.g. heart murmur detection. In practice, the envisioned use-case will determine the maximal compression ratio.

\section{Discussion}
The reduction in power consumption of wearable biosensor nodes has substantial implications, particularly for long-term deployments where battery life is a critical constraint. Optimizing energy efficiency enables these nodes to function continuously over extended periods, enhancing their feasibility for real-world applications in healthcare, sports, and human-computer interaction. A primary challenge lies in balancing reconstruction fidelity with power limitations. This trade-off is especially crucial in clinical and sports settings, where precise movement tracking and accurate biosignal acquisition are essential. The current firmware provides real-time adjustment of the compression ratio, to provide flexibility in the required analysis detail.

A significant advantage of reducing the data rate is the feasibility of utilizing low-power transmission technologies such as Bluetooth LE. In particular, applications such as phonocardiography (PCG) signal acquisition typically require data rates that exceed the theoretical limits of Bluetooth. However, by employing compressed sensing (CS), the use of Bluetooth instead of Wi\textendash Fi becomes viable. As demonstrated in Table \ref{table:powermeasurements}, this opportunity results in a substantial reduction in power consumption from \qty{25}{\milli\watt} to \qty{4.9}{\milli\watt}—approximately a factor of five—compared to Wi-Fi-based transmission. 

The achieved synchronization is single-sample accurate with a strong margin of $\qty{113,9}{\micro\second}$ to the highest sampling frequency utilized in the system ($\qty{8}{\kilo\hertz}$). This level of accuracy is not achievable with e.g. NTP, and justifies the added cost of the synchronization channel.

Furthermore, the speaker-microphone test setup confirmed the robustness of the reconstruction convolutional neural network (CNN) against variations in microphone transfer characteristics. This finding validates the system’s ability to maintain signal integrity without necessitating extensive retraining of the CNN on a large acquired dataset.

\section{Conclusion}
This work presents the development of a multimodal biosensor node that leverages compressive sensing for efficient signal acquisition and reconstruction. By integrating a CNN-based reconstruction framework, we demonstrate a significant reduction in power consumption while maintaining high-fidelity signal recovery. This approach is particularly beneficial for wearable biosensors, where energy efficiency directly impacts device longevity and usability in real-world scenarios.
Through experimental validation, including a speaker-microphone test setup, we have established the robustness of the reconstruction towards differences in microphone transfer characteristic. The proposed biosensor node holds promise for applications in clinical monitoring, rehabilitation, and sports performance analysis, where continuous and reliable physiological data collection is crucial.
Looking ahead, further optimizations will enhance the system’s efficiency and adaptability. Future work includes the implementation of low-power and sleep modes, an optimized network stack using WebSockets, implementation of the pseudo-random decimation in the PIO module instead of on the main core, and the exploration of alternative reconstruction architectures beyond CNNs. Additionally, expanding compressed sensing (CS) reconstruction to modalities such as EMG, ECG, and IMU data will broaden the scope of physiological monitoring applications. Evaluating alternative acoustic sensing methods, such as bell diaphragm-based audio capture, will further refine signal acquisition quality.
The advancements presented in this work have the potential to impact wearable health monitoring and physiological research significantly. By enabling low-power, high-fidelity biosignal processing, this system facilitates new possibilities in healthcare, wellness, and personal assistance technologies.

\bibliographystyle{IEEEtran}
\bibliography{allref}
\end{document}